\documentclass{emulateapj}
\usepackage{jjzhang}
\usepackage{xcolor}
\usepackage{graphicx,color,bm}
\usepackage{amsmath,amssymb}
\usepackage{epstopdf}
\usepackage{ulem}
\usepackage{array}
\usepackage{verbatim}
\usepackage[utf8]{inputenc}
\usepackage{rotating}
\newcolumntype{K}[1]{>{\centering\arraybackslash}m{#1}}
\usepackage{hyperref}

\begin{document}

\title{The first constraint from SDSS galaxy-galaxy weak lensing measurements on interacting dark energy models}

\author{Jiajun Zhang}
\affiliation{Department of Astronomy, School of Physics and Astronomy, Shanghai Jiao Tong University, Shanghai, 200240, China}
\email{liamzhang@sjtu.edu.cn}
\author{Rui An}
\affiliation{Department of Astronomy, School of Physics and Astronomy, Shanghai Jiao Tong University, Shanghai, 200240, China}
\affiliation{IFSA Collaborative Innovation Center, Shanghai Jiao Tong University, Shanghai, 200240, China}
\email{{an\_rui@sjtu.edu.cn}}
\author{Wentao Luo}
\affiliation{Department of Astronomy, School of Physics and Astronomy, Shanghai Jiao Tong University, Shanghai, 200240, China}
\affiliation{Kavli Institute for the Physics and Mathematics of the Universe (Kavli IPMU, WPI),
Tokyo Institutes for Advanced Study, The University of Tokyo, Chiba, 277-8582, Japan}
\author{Zhaozhou Li}
\affiliation{Department of Astronomy, School of Physics and Astronomy, Shanghai Jiao Tong University, Shanghai, 200240, China}
\author{Shihong Liao}
\affiliation{Key Laboratory for Computational Astrophysics, National Astronomical Observatories, Chinese Academy of Sciences, Beijing, 100012, China}
\author{Bin Wang}
\affiliation{Center for Gravitation and Cosmology, College of Physical Science and Technology, Yangzhou University, Yangzhou, 225009, China}
\affiliation{School of Aeronautics and Astronautics, Shanghai Jiao Tong University, Shanghai 200240, China}
\email{{wangb@yzu.edu.cn}}

\begin{abstract}
{We combine constraints from linear and nonlinear scales, for the first time, to study the interaction between dark matter and dark energy. We devise a novel N-body simulation pipeline for cosmological models beyond $\Lambda$CDM. This pipeline is fully self-consistent and opens a new window to study the nonlinear structure formation in general phenomenological interacting dark energy models. By comparing our simulation results with the SDSS galaxy-galaxy weak lensing measurements, we are able to constrain the strength of interaction between dark energy and dark matter. Compared with the previous studies using linear examinations, we point to plausible improvements on the constraints of interaction strength by using small scale information from weak lensing. This improvement is mostly due to the sensitivity of weak lensing measurements on nonlinear structure formation at low redshift. With this new pipeline, it is possible to look for smoking gun signatures of dark matter-dark energy interaction.}

\end{abstract}
\maketitle
\section{Introduction}
The current standard cosmological model, $\Lambda$CDM model, is widely accepted in explaining various astronomical observations \citep{Planck2016AA,Perlmutter1998Natur,Riess1998AJ,Perlmutter1999ApJ,Begeman1991MNRAS,Persic1996MNRAS,Chemin2011AJ}. However recently some observational tensions have been reported if the universe is described by the $\Lambda$CDM model. It was found that there is about $3\sigma$ mismatch for the Hubble constant inferred from the cosmic microwave background (CMB) measurements and from the direct local observations if the $\Lambda$CDM model is assumed \citep{Riess2011,Riess2016}. Besides, the Baryon Oscillation Spectroscopic Survey (BOSS) experiment showed that there is a $2.5\sigma$ deviation from the $\Lambda$CDM model in the measurement of the Hubble constant and angular distance at an average redshift $z=2.34$ \citep{Delubac2015}. Furthermore, a `substantial discordance' at the level of $2.3\sigma$ was obtained between the weak lensing data taken from a $450-\text{deg}^2$ observing field of the Kilo Degree Survey (KiDS) and the Planck 2015 CMB data \citep{Hildebrandt2017} if the $\Lambda$CDM model is supposed. Besides observational challenges, the $\Lambda$CDM model faces serious theoretical problems, such as the cosmological constant problem \citep{Weinberg1989} and the coincidence problem \citep{Zlatev1999}. This motivates us to find some more viable models to describe our universe. 

In the framework of Einstein gravity, nearly 95\% of the universe content is composed of dark matter(DM) and dark energy(DE). From the field theory point of view, it is a specific assumption that DM and DE live independently in the universe. More naturally we can consider some interactions between these two biggest components. The interaction between DM and DE has been discussed extensively in the literature, for a recent review please see \citet{Wang2016} and references therein. It is interesting to find that appropriate interaction between dark sectors can relieve discordances in observations as previously inferred from the $\Lambda$CDM model \citep{costa2017jcap,Ferrerira2017PRD,An2018JCAP}. Moreover, the coincidence problem can be alleviated if there is a proper interaction between DM and DE \citep{He2011PhRvD}.

The influence of interacting dark energy (IDE) models on the background dynamics and the linear perturbation evolutions in the universe has been studied extensively, see the review of \citet{Wang2016} and the references therein. In the nonlinear regime, N-body simulations are essential to understanding the structure formation and evolution. A preliminary attempt on the N-body simulation by considering quintessence DE interacting with DM was proposed in \citet{baldi2010mnras,baldi2011mnras}, where the initial condition in the simulation was naively taken from the $\Lambda$CDM model and the DE perturbation was not consistently computed at different scales and redshifts.  For general phenomenological IDE models, self-consistent N-body simulations are still lacking.

In this Letter, we devise a novel cosmological N-body simulation pipeline for cosmological models beyond $\Lambda$CDM. We consider self-consistent initial conditions for IDE models and include DE distributions from directly solving perturbation equations. We do not limit the DE in the quintessence region and consider general DE fluid phenomenologically interacting with DM.  We apply our simulation pipeline to four types of IDE models \citep{Wang2016,costa2017jcap}, and try to explore the physics in the structure formation when there are interactions between dark sectors.  
With this self-consistent and effective pipeline, we open a new window to precisely study the nonlinear structure formation in IDE models at low redshifts. This enables us to employ a new probe, weak lensing, to put further constraints on IDE models. 
For the first time, we use the galaxy-galaxy weak lensing measurements from the SDSS data \citep{luo2017apj,luo2018apj} to compare with our simulation results. 
We find that the improvement of the constraint for the interaction strength in some IDE models can reach up to $1250\%$. This shows the power of our cosmological N-body simulation pipeline in studying IDE models. 
With this tool, we can refine IDE models allowed by linear constraints \citep{costa2017jcap}. Finally, We are able to look for smoking gun signatures of interactions between the dark sectors, in the simulations constrained by current observations.

\section{Phenomenological Models}
The interaction between dark sectors is well motivated from field theory and is widely discussed in literatures, see recent review of \citet{Wang2016}. With the interaction between dark sectors, the background continuity equations of DM and DE obey
\begin{equation}
\label{eq:interaction}
\dot{\rho_c}+3H\rho_c=Q,
\dot{\rho_d}+3H(1+w_d)\rho_d=-Q.
\end{equation}
Here we focus on the commonly assumed phenomenological interaction form $Q=3\xi_1H\rho_c+3\xi_2H\rho_d$, where $\rho_c$ is the DM density, $\rho_d$ is the DE density and the dot denotes the derivative with respect to the conformal time, $H$ is the Hubble parameter and $w_d=p_d/\rho_d$ is the equation of state for DE. We do not limit DE to be a quintessence field with $w_d>-1$ \citep{baldi2011mnras.a}, but instead allow $w_d$ to be a free value either in the quintessence or the phantom regions.  $\xi_1$ and $\xi_2$ indicate the strength of interactions.

The linear evolutions of density and velocity perturbations for DM and DE were described in \citet{He2008,He2009,me-gadget}. In the subhorizon approximation, from linear level equations we can obtain the Poisson equation in the real space
\begin{equation}
\label{eq:poisson}
\nabla^2\Psi=-\dfrac{3}{2}H^2[\Omega_c\Delta_c+(1-\Omega_c)\Delta_d],
\end{equation}
where $\Delta_d$ ($\Delta_c$) is the density perturbation of DE (DM), and $\Omega_c$ is the background density ratio of DM. It is clear that with the interaction, the gravitational potential is modified.
The corresponding Euler equation in the real space reads
\begin{equation}
\label{eq:euler}
\nabla\dot{v_c}+[H+3H(\xi_1 +\dfrac{\xi_2}{r})]\nabla v_c=\nabla^2\Psi,
\end{equation}
where $r=\rho_c / \rho_d$. The coupling between dark sectors introduces an additional acceleration on DM particles at each time step in the simulation.
In the following we will concentrate our discussions on phenomenological IDE models listed in Tab.~\ref{tab.model} \citep{costa2017jcap}, which are natural Taylor expansions of the interaction kernel $Q$ into energy densities $\rho_c$ and $\rho_d$.
\begin{table}[ht]
\caption{Phenomenological IDE models \label{tab.model}}
\begin{center}
\begin{tabular}{ccc}
\hline
Model & $Q$ & $w$\\
\hline
I & $3\xi_{2}H\rho_{d}$ & $-1<w_d<-1/3$ \\
II & $3\xi_{2}H\rho_{d}$ & $w_d<-1$ \\
III & $3\xi_{1}H\rho_{c}$ & $w_d<-1$ \\
IV & $3\xi H(\rho_{c}+\rho_{d})$ & $w_d<-1$ \\
\hline
\end{tabular}
\end{center}
\end{table}
With Planck 2015, Type Ia supernovae, baryon acoustic oscillations, and the Hubble constant observation,  tight constraints on $\xi_1$ for Model III and $\xi$ for Model IV were obtained in \citet{costa2017jcap}. However, for Models I and II, the obtained constraints on the strengths of couplings are loose \citep{costa2017jcap}. This is well expected because in Models I and II, the interaction is proportional to the energy density of DE, which was sub-dominant when CMB was produced. It is more reasonable to expect that the observations at low redshifts shall provide tighter constraints on Models I and II, especially the small-scale structure information.  For this purpose we resort to using N-body simulations to make accurate analysis.

\section{Simulation Pipeline}
Since IDE models are different from the $\Lambda$CDM model in every relevant equation, it is naive to count on empirical fits to the $\Lambda$CDM model, e.g. the halofit, to understand the physics in the nonlinear structure formation. We require a new N-body simulation pipeline to understand the structure developed in IDE models. There are four modifications we have considered in devising the new pipeline compared to the standard $\Lambda$CDM model. Firstly, the pre-initial condition is generated by the Capacity Constrained Voronoi Tessellation (CCVT) method \citep{liao2018ccvt}, instead of the classically used glass or grid. This makes sure that our pre-initial condition is free of Poisson equation at all, generating geometrically equilibrium state of particle distribution. We have tested that using CCVT, grid or glass makes negligible differences for the pre-initial condition in simulations. The choice of CCVT is mainly because of self-consistency consideration, rather than the accuracy consideration. Secondly, the initial matter power spectrum is generated by our modified CAMB \citep{camb,costa2017jcap} with the coupling between dark sectors, which is different from the $\Lambda$CDM model. Thirdly, the perturbations of the particle distribution are calculated by using 2LPTic \citep{2lptic}, which is properly modified to be consistent with our models. Fourthly, the N-body simulation code is also heavily modified for consistency.

Instead of treating the DE perturbation as a constant excess of gravity at all scales and redshifts  \citep{baldi2010mnras,baldi2011mnras}, we include the DE perturbation self-consistently as a function of scale ($k$) and redshift ($z$) by solving perturbation equations from the modified CAMB \citep{costa2017jcap}. We also modify the N-body simulation code Gadget2\citep{gadget2} into ME-Gadget. Technical details can be found in \citet{me-gadget}. We find that ME-Gadget is as efficient as the original Gadget2 code, and the testing results are consistent with \citet{baldi2011mnras} by using their models. Our convergence test results also show that our code can reach $5\%$ accuracy as $k$ approaching the Nyquist limit for the nonlinear matter power spectrum at $z=0$. We would like to emphasize that our N-body simulation pipeline is fully self-consistent, accurate and efficient enough for general phenomenological IDE models.
The simulation parameters we use are shown in Tab.~\ref{tab.parameters}, which were constrained from the combination of Planck 2015, Type Ia supernovae, baryon acoustic oscillations, and the Hubble constant observation datasets (PBSH in short hereafter) \citep{costa2017jcap}.  We use a comoving box size of $400h^{-1}$Mpc and $256^3$ particles in our computations for the matter power spectrum. A comoving box size of $400h^{-1}$Mpc and $512^3$ particles is used for the measurements of galaxy-galaxy lensing signals in the simulations. 
\begin{table}
\caption{Cosmological parameters}
\label{tab.parameters}
\begin{tabular}{cccccc}
\hline
Parameter & IDE\_I & IDE\_II & IDE\_III & IDE\_IV & $\Lambda$CDM\\
\hline
$\Omega_b h^2$ & 0.02223 & 0.02224 & 0.02228 &  0.02228 & 0.02225\\
$\Omega_c h^2$ & 0.0792  & 0.1351  & 0.1216  & 0.1218 & 0.1198\\
$100\theta_{MC}$ & 1.043 & 1.04 & 1.041 & 1.041 & 1.04077\\
$\tau$  & 0.08204 & 0.081 & 0.07728 & 0.07709 & 0.079\\
${\rm{ln}}(10^{10} A_s)$ & 3.099 & 3.097  & 3.088  & 3.087 & 3.094\\
$n_s$ & 0.9645 & 0.9643 & 0.9624 & 0.9624 & 0.9645\\
$w$ & -0.9191 & -1.088 & -1.104  & -1.105 & -1 \\
$\xi_1$ & -- & -- & 0.0007127 & 0.000735 & -- \\
$\xi_2$ & -0.1107 & 0.05219 & -- & 0.000735 & --\\
\hline
$H_0$ & 68.18 & 68.35 & 68.91 & 68.88 &67.27 \\
$\Omega_m$ & 0.2204 & 0.3384 & 0.3045 & 0.3053 & 0.3156\\
\hline
\end{tabular}
\end{table}

\section{Matter Power Spectrum}
\begin{figure}
\includegraphics[width=0.5\textwidth]{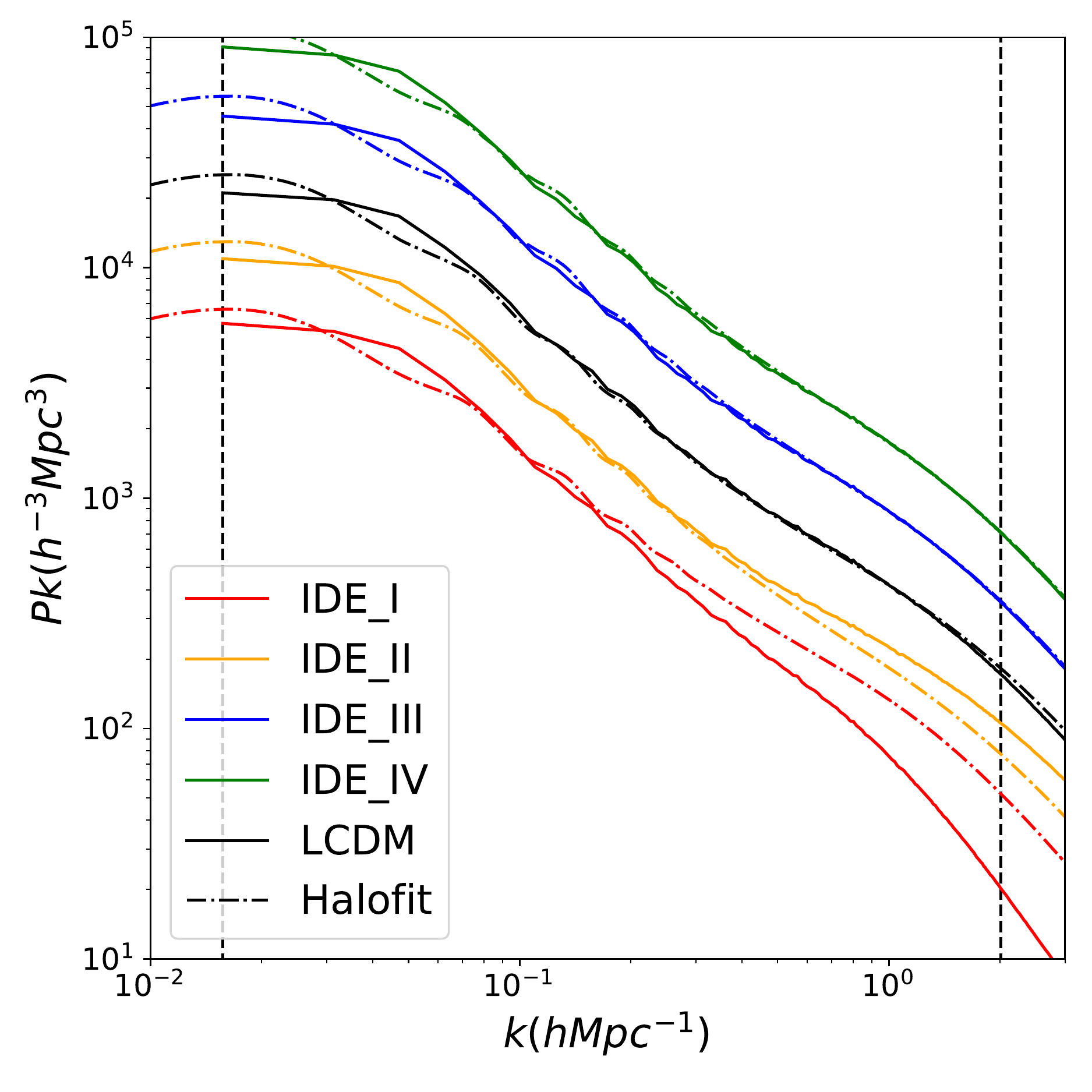}
\caption{Different colors show matter power spectra of different models at $z=0$. Solid lines are measured from N-body simulations while the dash-dotted lines are predicted nonlinear matter power spectra using halofit \citep{takahashi2012halofit}. The vertical dashed lines show the credible range of matter power spectrum, the box size limit on the left and the Nyquist limit on the right. We have rescaled IDE\_I (IDE\_II, IDE\_III, IDE\_IV) by a factor of $\dfrac{1}{4}$ ($\dfrac{1}{2}, 2, 4$) for a better illustration. All the models are identical with $\Lambda$CDM (LCDM) at large scales ($k<0.1hMpc^{-1}$). It is clear that halofit is not correct for IDE\_I and IDE\_II.}\label{fig:pk}
\end{figure}

The matter power spectrum is used to quantify the large-scale structures. The linear evolution of the matter power spectrum can be simply calculated by linear growth theory. People usually use halofit \citep{takahashi2012halofit} to estimate the non-linear matter power spectrum at low redshifts. However, since halofit is an empirical fit to $\Lambda$CDM N-body simulations, it is not appropriate to use it to describe IDE models. To make it clear, we compare the measured matter power spectrum from our N-body simulations with the prediction of halofit in Fig.~\ref{fig:pk}. The matter power spectrum is computed using the ComputePk code \citep{computepk}. We find that halofit can marginally be used to describe the nonlinear matter power spectra at $z=0$ for IDE\_III and IDE\_IV models, although it is not exactly consistent with that from N-body simulations. This is because the strengths of the interaction in these two models are quite small  ($\sim0.0007$) so that the deviations from the  $\Lambda$CDM model are negligible. However, for models IDE\_I and IDE\_II, it is clear that halofit cannot give the true matter power spectrum, especially at small scales, because their interactions are relatively large ($|\xi_2|>0.05$) which cause large deviations from the standard $\Lambda$CDM model. The empirical fit to $\Lambda$CDM is no longer appropriate in these cases to describe the nonlinear structure, and the appropriate N-body simulations pipeline is called for.

\section{Galaxy-Galaxy Lensing}
The galaxy images are distorted by the foreground gravitational potential,
which is known as gravitational lensing. Assuming an isotropic distribution
of both galaxy shape and orientation, any non-zero residual can be considered
as such effect, a.k.a. tangential shear $\gamma_T$. In galaxy-galaxy lensing,
the signal is usually interpreted as the combination of $\gamma_T$ and
the geometry of a lensing system, referring to the critical density
$\Sigma_{crit}(z_l,z_s)=\frac{c^2}{4\pi G}\frac{D_s}{D_{ls}D_l}$, where $z_l, z_s$ denote the redshifts of the lens and the source, $D_l$,
$D_s$ and $D_{ls}$ are the angular diameter distances of the lens, source
galaxy and the difference between them. The signal measured
from galaxy-galaxy lensing actually reflects the differential change of 2D surface
density-Excess Surface Density(ESD),
\begin{equation}
\Delta\Sigma(R)=\Sigma(\leq R)-\Sigma(R)=\gamma_t \Sigma_{crit}(z_l,z_s),
\end{equation}
here $\Sigma(\leq R)$ is the average surface density inside the projected distance $R$ and $\Sigma(R)$ is the surface density at the projected distance $R$. This signal is multiplied by a factor, a.k.a boost factor to correct for the contamination by galaxies associated with lens galaxy. This factor is calculated following \cite{Mandelbaum2005}
\begin{equation}
    B(R)=\frac{n(R)}{n_{rand}(R)},
\end{equation}
where $n(R)$ and $n_{rand}(R)$ are the background numbers around lens sample at radius R. 

We use the shear catalog from \citet{luo2017apj}, which is based on the SDSS DR7 image data. For groups of galaxies, we employ the catalog from \citet{yang2007apj} to identify the lens systems. Following the galaxy-galaxy lensing measurement procedure in \citet{luo2018apj},  we select the most luminous 3660 galaxy groups in the group catalog from redshift 0.01-0.2 as the lens. To estimate the abundance, we calculate the comoving volume of the SDSS DR7 north cap between redshift 0.01 and 0.2, labeled as $V_{com}$, with completeness considered. The completeness of each galaxy is given by the NYU-VAGC \citep{Blanton2005}. The number of halos with same abundance in the simulation is then estimated as $3660\times400^3/V_{com}\approx1771$.  We stack the tangential shear of these 3660 lens systems to measure the ESD. Taking the halos of the same abundance as that in the observation in our N-body simulations, we select the most massive 1771 halos and stack their particles to measure the ESD.  The systematics introduced by photometric redshift is about $2.5-2.7\%$ estimated based on EQ.(24) in \citet{Mandelbaum2005}.  About 5\% of the galaxies are satellites according to \citet{Cacciato2009}, the contribution to the ESD is roughly about 4\% at 300$h^{-1}kpc$ to 10\% contribution to the maximum at 1000$h^{-1}kpc$. These uncertainties and bias are neglected for the analysis below because they are too small to affect our final results.

\begin{figure}
\includegraphics[width=0.5\textwidth]{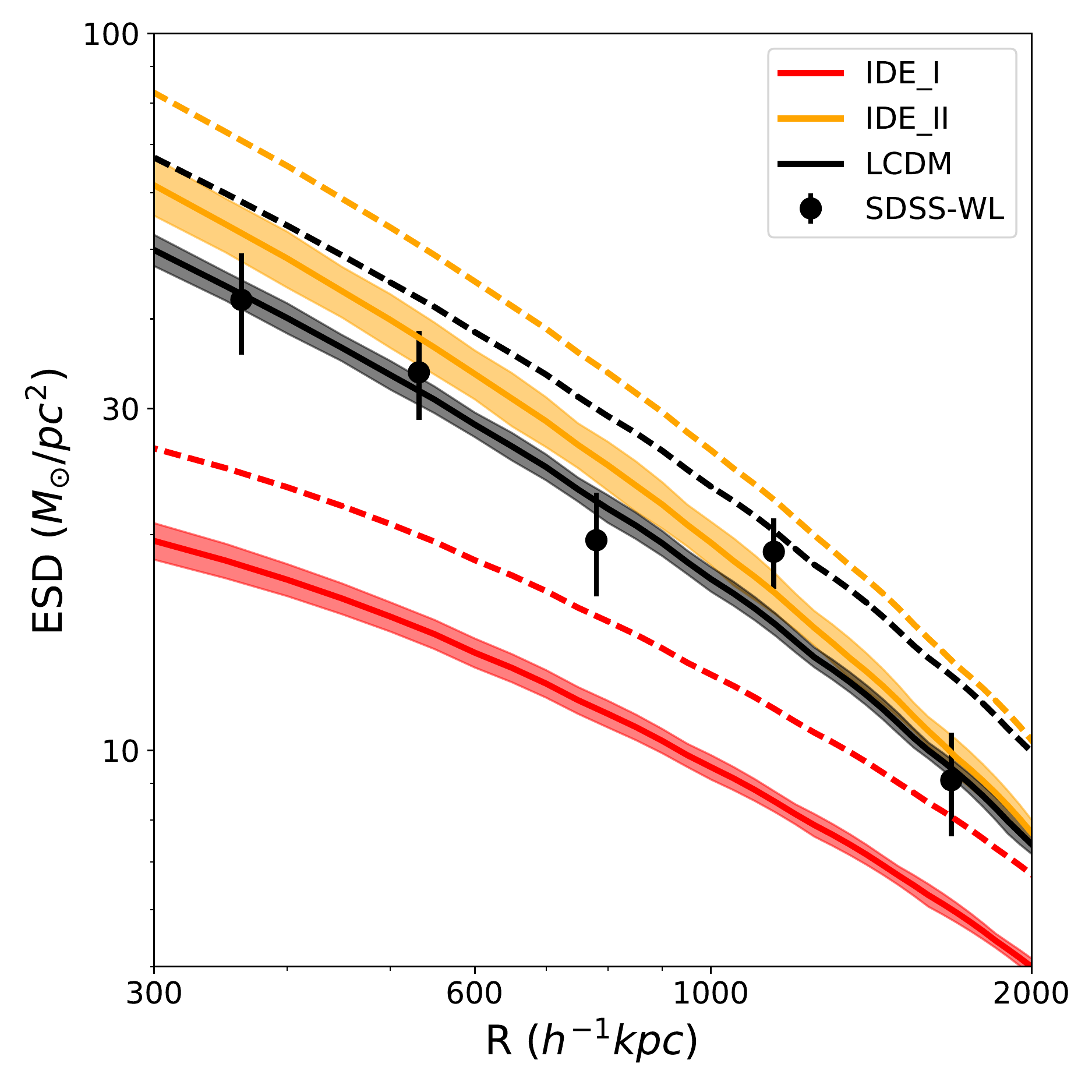}
\caption{The Excess Surface Density (ESD) measured from IDE\_I, IDE\_II and $\Lambda$CDM (LCDM) model simulations is shown in red, orange and black respectively. The shaded regions show the ESD range between $z=0.1$ and $z=0.2$, which illustrate the redshift uncertainty. The solid (dashed) lines show the results with (without) the Eddington bias corrections. The $\Lambda$CDM model is clearly more favored by the SDSS galaxy-galaxy weak lensing data (SDSS-WL) than IDE\_I and IDE\_II. Because IDE\_III and IDE\_IV results are almost identical to $\Lambda$CDM, we hide them for a better illustration. \vspace{2mm}}\label{fig:esd}

\end{figure}

The weak lensing measurements and simulation predictions of each cosmological model are shown in Fig.~\ref{fig:esd}. By comparing the ESD curves from the $256^3$ and $512^3$ simulations, we find that the $256^3$ results converge to the $512^3$ ones with a level of $<5\%$ at $r>600$ kpc for all three cosmological models (LCDM, IDE\_I and IDE\_II). Therefore, we expect that the ESD curves from the $512^3$ simulations shown in Fig.2 should have a convergence level of $<5\%$ at $r>300$ kpc for all models. We find that the measured data points are systematically lower than the prediction from the $\Lambda$CDM model shown in black dashed line, which is mainly due to the Eddington bias \citep{luo2018apj}. The Eddington bias comes from the incorrect estimation of the halo mass using the galaxy luminosity or other indicators. The incorrect estimation will mistakenly identify lower mass halos as higher mass halos, thus contaminate the ESD signal. We corrected the Eddington bias by assuming a 0.3 dex scatter in mass-luminosity relation following \citet{luo2018apj}, shown as the solid lines. We have tested that the Eddington bias introduced in \citet{luo2018apj} is similar for IDE models by using the halo catalogs from our simulations. The shaded area represents the dispersion due to finite width of redshift bin. The groups of galaxies we selected locate at different redshifts, central at $z=0.15$ (range $0.01<z<0.2$). Thus, the uncertainty due to the redshift difference was also taken into account in our analysis. We estimate the redshift bin by measuring the ESD signal from simulation snapshots at $z=0.1$ and $z=0.2$ separately. The solid lines show the central value of the shaded area with the same colors. Even with such conservative treatments, it is still quite clear that IDE\_I and IDE\_II are not favored by the SDSS galaxy-galaxy weak lensing data, even though these two IDE models are well constrained by PBSH. Therefore, tight constraints from comparing our simulations with observational galaxy-galaxy lensing signals are expected.

\section{Constraints}
\begin{figure}
\includegraphics[width=0.5\textwidth]{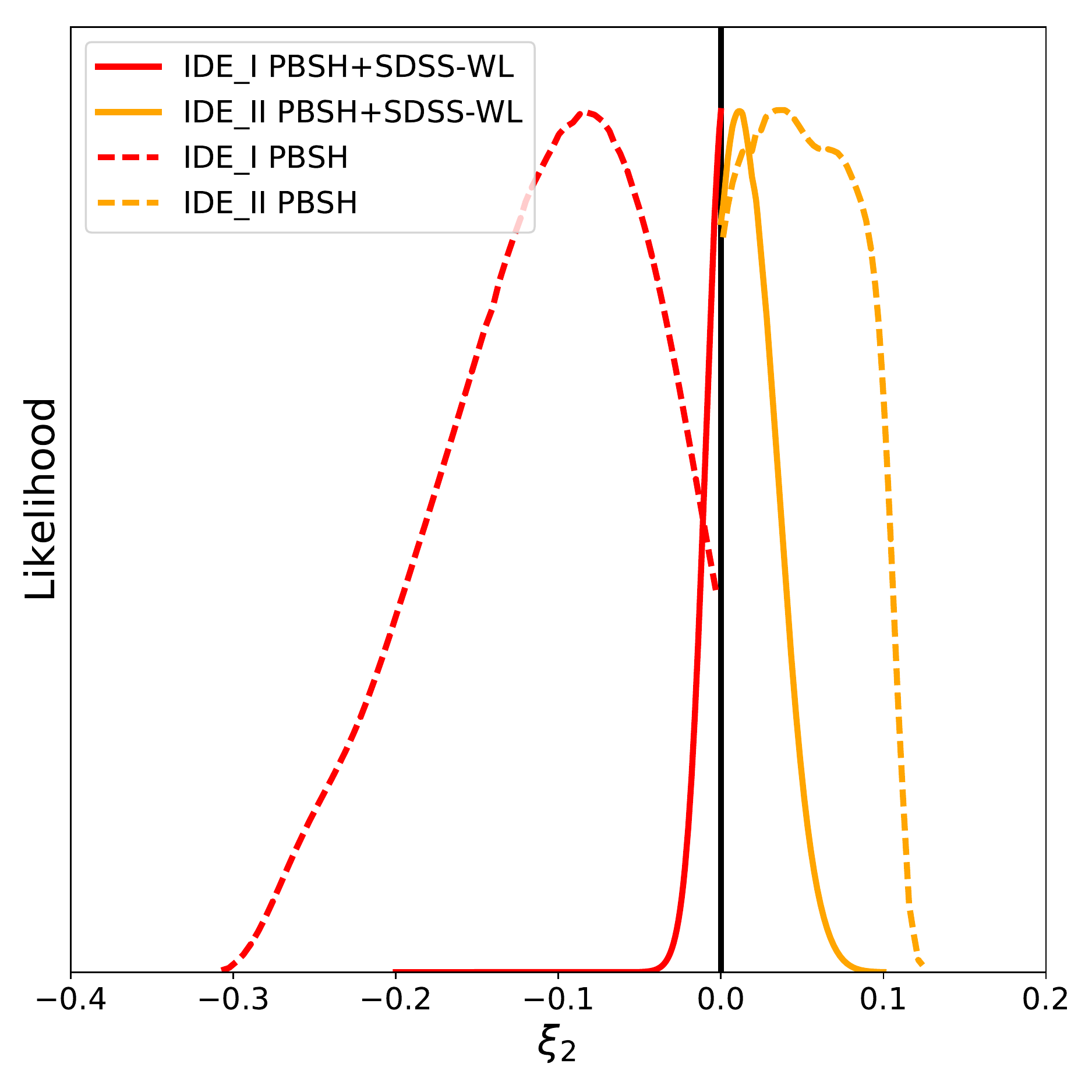}
\caption{The constraints of $\xi_2$ are shown in red (orange) lines for IDE\_I (IDE\_II). The dashed lines show the constraints from Planck 2015, Baryonic Acoustic Oscillation, Supernovae Type Ia and H0 observations, labeled PBSH in short. The solid lines show the combined constraints from PBSH and SDSS galaxy-galaxy lensing. The improvement for IDE\_I and IDE\_II is huge. The one sigma lower bound for IDE\_I is $\xi_2=-0.0105$, while the one sigma upper bound for IDE\_II is $\xi_2=0.0286$. Comparing to PBSH only, the improvement of constraints is $\sim1250\%$ for IDE\_I and $\sim260\%$ for IDE\_II.}\label{fig:likelihood}
\end{figure}

We estimate the constraints from galaxy-galaxy lensing signals by assuming that the ESD signal deviation from the $\Lambda$CDM model in logarithmic space is linearly proportional to the interaction strength. We have tested that the above assumption is not significantly affected by the choice of logarithmic space or linear space. We have also tested that such an assumption is reasonably accurate using multiple simulations with different parameters. The likelihood is constructed as
\begin{equation}
L = \exp\left\{-\frac{1}{2}\sum_i\dfrac{\left[\Delta\Sigma(R_i)_{sim}-\Delta\Sigma(R_i)_{obs}\right]^2}{\sigma_z^2+\sigma_{obs}^2}\right\}.
\end{equation}
Here $R_i$ denotes the measured five data points, $\sigma_z=0.288$ times the width of the shaded area, representing the uncertainty due to the finite width of the redshift bin, and $\sigma_{obs}$ is the error estimated from the lensing signal. We show the likelihood from our comparison in Fig.~\ref{fig:likelihood}. Comparing to the linear constraints given by \citet{costa2017jcap} shown in dashed lines, the constraints from our SDSS galaxy-galaxy weak lensing (SDSS-WL) are clearly tighter for models IDE\_I and IDE\_II.  The improvements of constraints for models IDE\_III and IDE\_IV are negligible and we do not show here. The joint likelihood of PBSH and SDSS-WL is about 1250\% tighter than PBSH alone for IDE\_I. The best-fitted $\xi_2$ for IDE\_I is $\xi_2=0$, and the best-fitted $\xi_2$ for IDE\_II becomes $\xi_2=0.012$.

\section{Conclusion}
We have successfully devised a self-consistent N-body simulation pipeline to examine the influence of the interaction between dark sectors in structure formation at low redshifts. This formalism is appropriate to general IDE models and efficient in examining the signature of the interaction. With this tool at hand, we do not need to blindly count on halofit, which is an empirical fit to the $\Lambda$CDM model, to disclose nonlinear structures.

Considering that interactions in IDE\_I and IDE\_II models are proportional to the energy density of DE, which is sub-dominant at high redshifts, it is natural to find that the constraints of these interactions from PBSH are loose. With the self-consistent N-body simulation pipeline however, we can examine these two models more carefully by using the nonlinear low redshift observations, such as the SDSS galaxy-galaxy weak lensing. It is interesting to find that our first try of the pipeline can obtain up to $1250\%$ improvement of the interaction strength constraint for the IDE\_I model. Combining PBSH and SDSS galaxy-galaxy weak lensing measurements, we find the constraint of the interaction strength $\xi_2=0^{+0.0}_{-0.0105}$ for the IDE\_I model. The $\Lambda$CDM model is still favored. For the IDE\_II model, combing PBSH and SDSS galaxy-galaxy weak lensing datasets we obtain $\xi_2=0.0120^{+0.0166}_{-0.012}$, which is also improved significantly by including the nonlinear structure information. For IDE models III and IV, the galaxy-galaxy lensing constraints by employing N-body simulations do not improve much of the constraints if we compare with the linear PBSH results. It is interesting that our pipeline is effective in disclosing physics in the structure formation when there is coupling between dark sectors and it can also help to refine IDE models. We would like to address that the likelihood and improvement from SDSS-WL is only a rough estimation about the parameter rather than a complete constraint. It is useful to guide our future study.

By combining the linear and nonlinear scale information, we can, not only constrain the interaction strength between dark sectors with much higher precision, but also perform the simulations constrained by the current observations. In such simulations, we can look for smoking gun signatures of the dark matter dark energy interactions. These signatures can be directly tested by the observations in the future.

\textit{Acknowledgement}.--
J.Z acknowledges the support from China Postdoctoral Science Foundation 2018M632097.
W.L acknowledges the support from NSFC 11503064 and Shanghai Jiao Tong University
and University of Michigan Joint Fundation (AF0720054). The work of B. W was partially supported by NNSFC.
This work was supported in part by World Premier International Research Center Initiative (WPI Initiative), Japan. We also thank Surhud More from IUCAA, India
and Feng Shi from KASI Korea for useful discussion.
\bibliographystyle{plain}
\expandafter\ifx\csname natexlab\endcsname\relax\def\natexlab#1{#1}\fi
\providecommand{\url}[1]{\href{#1}{#1}}
\providecommand{\dodoi}[1]{doi:~\href{http://doi.org/#1}{\nolinkurl{#1}}}
\providecommand{\doeprint}[1]{\href{http://ascl.net/#1}{\nolinkurl{http://ascl.net/#1}}}
\providecommand{\doarXiv}[1]{\href{https://arxiv.org/abs/#1}{\nolinkurl{https://arxiv.org/abs/#1}}}

\end{document}